\documentclass[]{article}
\newcommand{\bb}{\begin{eqnarray}}
\newcommand{\ee}{\end{eqnarray}}
\begin{document}

\title{
Asymptotically non-flat rotating dilaton black holes} 

\author{Tanwi Ghosh and P Mitra\thanks{e-mail tanwi@theory.saha.ernet.in, 
mitra@theory.saha.ernet.in}\\
Theory Division\\
Saha Institute of Nuclear Physics\\
1/AF, Bidhannagar, Calcutta 700 064, India
}

\date{\tt gr-qc/0212057}
\maketitle

\begin{abstract}
Dilaton black hole solutions which are neither
asymptotically flat nor (anti)-de Sitter but reduce to asymptotically flat
solutions in some special limits have been known for a
Liouville type dilatonic potential. It is shown how,
by solving a pair of coupled differential equations,
infinitesimally small angular momentum can be added to these static
solutions to produce rotating black hole solutions. 
\end{abstract}

%\pacs{04.70.Bw}

\section{Introduction}
A scalar field called the dilaton occurs in the
low energy limit of string theory. The Einstein action can be modified
to include a dilaton term in addition to the familiar Maxwell term,
which again gets modified by a dilaton coupling. 
Exact solutions for charged dilaton black holes have been constructed 
by many authors. 
The dilaton changes the causal structure of the black hole and
leads to curvature singularities at finite radii [1 - 7].  
These black holes are asymptotically flat. In the presence of
a Liouville potential generalization of a cosmological constant, a class of
electrically (or magnetically) charged static spherically symmetric
dilaton black hole solutions have also been discovered [8]  
with metrics which are asymptotically neither flat nor (anti)-de Sitter.  

The action is given by
\bb
S=\int d^4x\sqrt{-g}[R-2(\nabla\phi)^2-V(\phi)-e^{-2\alpha\phi}F^2],
\ee
where $R$ is the scalar curvature, $F^2=F_{\mu\nu}F^{\mu\nu}$ is the
usual Maxwell contribution, $V(\phi)=2\Lambda e^{2b\phi}$, a Liouville
potential involving a dilaton field $\phi$, $\alpha$  determining the 
strength of coupling between the electromagnetic field and $\phi$. 
$\Lambda$ is of course the cosmological constant
to which the potential reduces for vanishing dilaton field, but
in the presence of a non-trivial dilaton field, the spacetime may not
behave as either de Sitter ($\Lambda> 0$) or anti-de Sitter ($\Lambda < 0$).  
In fact, it has been
shown that with the exception of a pure cosmological constant potential,
where $b = 0$, no  asymptotically de Sitter or
asymptotically anti-de Sitter static spherically symmetric solution to
the field equations associated with $V(\phi) = 2\Lambda e^{2b\phi}$ exist [9]. 

The exact solutions mentioned above are all static.
Unfortunately, exact rotating solutions to the Einstein equation coupled to 
matter fields are difficult to find except in a limited number of cases.
An exact solution for a rotating black hole with a
special dilaton coupling was derived using the inverse scattering method [10].  
For general dilaton coupling, the properties of asymptotically flat
charged dilaton black holes only with infinitesimally small angular momentum 
[11 - 12] or small charge [13] have been investigated. 
Our aim here is to investigate the possibility in the class of asymptotically
non-flat solutions [8] of adding an infinitesimally small rotation.

\section{Field equations and solutions}
Varying the action with respect to the metric and the Maxwell 
and dilaton fields respectively leads to
\bb
R_{ab}&=&2\partial_a\phi\partial_b\phi+\frac12 g_{ab}V+2e^{-2\alpha\phi}
[F_{ac}{F_b}^c-\frac14 g_{ab}F_{cd}F^{cd}],\\
0&=&\partial_a[\sqrt{-g}e^{-2\alpha\phi}F^{ab}],\\
\nabla^2\phi&=&\frac14{\partial V\over\partial\phi}-
\frac\alpha2 e^{-2\alpha\phi}
F_{cd}F^{cd}.
\ee

We  take the metric to be of the form 
\bb
ds^2=-u(r)dt^2+{dr^2\over u(r)}+h(r)d\Omega^2-2af(r)\sin^2\theta dtd\phi,
\ee
where $a$ is an angular momentum parameter.  
This metric is static and spherically symmetric when $a=0$. If $a$ is small,
one can expect to find solutions with $u$ still a function of $r$ alone.
The Maxwell equations can be integrated
to give 
\bb
-A_0'=F_{01}={Qe^{2\alpha\phi}\over h},
\ee
where $Q$, an integration constant, is the electric charge, but in 
the presence of rotation, there will in general also be a vector potential
\bb
A_\phi=aQc(r)\sin^2\theta,
\ee
where $c(r)$ satisfies
\bb
Q(uc'e^{-2\alpha\phi})'-{2Qce^{-2\alpha\phi}\over h}=-Q\bigg({f\over h}\bigg)'.
\label{c}\ee
A common factor $Q$ has been retained in all terms of this equation
in order to emphasize that the equation disappears when $Q=0$.
With the above metric and Maxwell fields, the remaining
equations of motion reduce for small $a$ to
\bb 
{1\over h}(hu\phi')'&=&b\Lambda e^{2b\phi}+
{\alpha Q^2e^{2\alpha\phi}\over h^2},\\
\bigg({h'^2\over 2h^2}-{h''\over h}\bigg)&=&2\phi'^2,\\
u''-{uh''\over h}+{1\over h}&=&{2Q^2e^{2\alpha\phi}\over h^2},\\
{1\over h}(uh')'&=&{2\over h}-V-{2e^{2\alpha\phi}Q^2\over h^2},\\
{f''\over f}&=&{h''\over h}-{4Q^2c'\over fh}.\label{f}
\ee

The equations involving $f,c$, namely the previous equation (\ref{f}) 
and the equation (\ref{c}) 
arising from the presence of $A_\phi$, appear only when $a\neq 0$, while
the other equations were there also in the static, spherically symmetric case.
Solutions of those equations for $h, u, \phi$ with unusual asymptotic
behaviour were given to ${\cal O}(a^0)$ in [8]. 

\subsection{First black hole}
The first non-rotating solution  [8] is
\bb
u(r)&=&r^{2\alpha^2\over 1+\alpha^2}\bigg({1+\alpha^2\over 1-\alpha^2}
-{2(1+\alpha^2)M\over r}+{Q^2(1+\alpha^2)\over r^2}\exp 2\alpha\phi_0\bigg),\\
h(r)&=&r^{2\over 1+\alpha^2},\\
\phi(r)&=&\phi_0-\log r\frac{\alpha}{1+\alpha^2},
\ee
with $\Lambda=-{\alpha^2\exp -2b\phi_0\over 1-\alpha^2}, b=\frac{1}{\alpha}$.
Horizons are located at 
\[ r_\pm=(1-\alpha^2)M\bigg(1\pm(1-{Q^2
\exp 2\alpha\phi_0\over M^2(1-\alpha^2)})^{-1/2}\bigg)\]
if $\alpha^2<1$.
Rotation of test particles can be studied: the velocity
for stable circular motion with $\theta=\pi/2$ 
turns out to be $\alpha$ independent of $r$ for large $r$.

We are interested here in rotating solutions.
To ${\cal O}(a)$, we can get a solution of the full set of equations 
with $u,h,\phi$ unaltered
and supplemented by the solutions of the new equations
(\ref{c}, \ref{f}) for $c,f$. 
These two coupled equations admit different solutions:
\bb
{\rm I:}&& c(r)=f(r)=0,\\
{\rm II:}&& c(r)=0,f(r)\propto h(r),\\
{\rm III:}&& c(r)=\frac1r,\nonumber\\
&&f(r)=(1+\alpha^2)\bigg(-2M(\exp -2\alpha\phi_0)r^{\alpha^2-1\over 1+\alpha^2}
+Q^2r^{-{2\over 1+\alpha^2}}\bigg).
\ee

Solution I is equivalent to the non-rotating solution [8], but it satisfies the
new equations (\ref{f}, \ref{c}) involving $f(r),c(r)$ in addition 
to the other equations
which do not involve these functions. Solution II involves a change
of the metric from the non-rotating form
without any change of the Maxwell field and follows from the general
structure of the new equations for $f(r),c(r)$. This may be surprising at 
first sight because a rotation enters
the metric without any rotation in the charge:
this is possible because the function $f(r)$ does not obey conventional
boundary conditions for large $r$ and in fact increases with $r$.
The solution III decreases with increase of $r$ for $\alpha^2<1$. 
In general, this solution is not asymptotically flat or (anti)-de Sitter. 
The $\alpha\to\infty$ limit produces de Sitter behaviour for zero $M,Q$.
In the limit $\alpha\to 0$, the non-rotating version of the solution [8]
reduces to the Reissner - Nordstr\"{o}m form, and the rotating version 
III above reproduces the Kerr-Newman
modification thereof for small $a$, {\it viz.} 
\bb
f=-{2M\over r}+{Q^2\over r^2}.
\ee

The value of the gyromagnetic ratio is known to be 2 in this limiting 
Kerr-Newman case. It is of interest to examine it in the more general
situation when $\alpha\neq 0$. In the absence of asymptotic flatness,
new definitions of angular momentum and mass are required, but it turns
out that the ratio $J/M$ can be defined unambiguously. $g_{t\phi}$, which
is usually of the form $\sin^2\theta/r$, is in the present situation 
of the form $\sin^2\theta 
r^{\alpha^2-1\over \alpha^2+1}$, but the first non-leading term in $g_{tt}$
has the same power of $r$, from which one can see that
\bb
\bigg({J\over M}\bigg)|_{phys}=-ae^{-2\alpha\phi_0}.
\ee
Since $\partial_r A_\phi=-aQ\sin^2\theta/r^2$ (the magnetic field
is $1/r\sin\theta $ times this expression), the magnetic dipole
moment is
\bb
\mu|_{phys}=-aQ,
\ee
and since $F_{01}=Qe^{2\alpha\phi}/h=Qe^{2\alpha\phi_0}/r^2,$
\bb
Q|_{phys}=Qe^{2\alpha\phi_0}.
\ee     
Thus the gyromagnetic ratio is
\bb
g=2{\mu|_{phys}\over Q|_{phys}}/\bigg({J\over M}\bigg)|_{phys}=2,
\ee
independent of $\alpha$.

\subsection{Second black hole}
The second non-rotating solution [8] is 
\bb
u(r)&=&r^{2\over 1+\alpha^2}\bigg({1+\alpha^2\over 1-\alpha^2}(-1
+2Q^2\exp 2\alpha\phi_0)-{2M(1+\alpha^2)\over \alpha^2r}\bigg),\\
h(r)&=&r^{\frac{2\alpha^2}{1+\alpha^2}},\\
\phi(r)&=&\phi_0+\log r\frac{\alpha}{1+\alpha^2},,
\ee
with $\Lambda={\exp 2\alpha\phi_0\over 1-\alpha^2}
-{(1+\alpha^2)Q^2\exp 4\alpha\phi_0
\over 1-\alpha^2}, b=-\alpha$.
A single horizon is located at $r_h={2(1-\alpha^2)M \over (-1+2Q^2
\exp 2\alpha\phi_0)\alpha^2}$, when this is positive.
The velocity for stable circular motion with $\theta=\pi/2$ turns out in 
this case to be $1/\alpha$ independent of $r$ for large $r$.

We are interested in finding the rotating version of this static solution,
that is to say, in solving the corresponding coupled equations for $f,c$. 
As in the previous case, there exists the solution I ($c(r)=f(r)=0$) 
which is a trivial extension of the non-rotating solution to the equations
including rotation. The unusual solution
II ($c(r)=0,f(r)\propto h(r)$) where there is a change in the metric
without any change in the electromagnetic field also
exists as before. Again, this $f(r)$ increases with $r$. 
In addition, there are solutions like
\bb
{\rm III:}&&c(r)=r^{4\alpha^2\over 1+\alpha^2},\nonumber\\
&&f(r)= \exp -2\alpha\phi_0 \bigg( -{4\alpha^2\over 1+\alpha^2}
{u(r)c(r)\over r}+2 r^{1+3\alpha^2\over 1+\alpha^2}\bigg),
\ee
for which  $Q,\alpha$ have to be related to $\phi_0$ by 
\bb
2Q^2\exp 2\alpha\phi_0={\alpha^2+1\over 3\alpha^2-1}.
\ee
$\Lambda$ can be made to vanish in this condition with  $\alpha^2=$1 or 3.
This solution too shows an increase with increase of $r$.
Another solution of this kind is
\bb
{\rm IV:}&&c(r)=r^2+{2M(\alpha^2-1)\over\alpha^2}r,\nonumber\\
&&f(r)=(1+\alpha^2)\exp -2\alpha\phi_0 \bigg[-2{2-\alpha^2\over 3-\alpha^2}
r^{3+\alpha^2\over 1+\alpha^2}+\nonumber\\
&&2M{2-\alpha^2\over \alpha^2}r^{2\over 1+\alpha^2}
+4M^2{\alpha^2-1\over \alpha^4}r^{1-\alpha^2\over 1+\alpha^2}\bigg],
\ee
for which it is necessary that  $Q,\alpha$ are related to $\phi_0$ by 
\bb
2Q^2\exp 2\alpha\phi_0=2-\alpha^2.
\ee
$\Lambda$ can be made to vanish in this condition only for $\alpha^2=$1.

To obtain functions which decrease with increasing $r$,
one has to consider {\it asymptotic solutions} like
\bb
{\rm V:}&&c(r)=r^p, \nonumber\\
&&f(r)=-{4pQ^2\over p(p+{1-3\alpha^2\over 1+\alpha^2})}
r^{p+{1-\alpha^2\over 1+\alpha^2}}+Cr^{1-\alpha^2\over 1+\alpha^2},
\ee
with $C$ an arbitrary constant for $p>0$ and $C=0$ for $p<0$. 
This satisfies the differential equation for $c(r)$ 
{\it asymptotically} for large $r$ 
and the differential equation for $f(r)$ holds exactly provided that
$p,Q,\alpha,\phi_0$ are related by 
\bb
2Q^2\exp 2\alpha\phi_0={-{1\over 1-\alpha^2}-{2\over p(1-3\alpha^2)+
p^2(1+\alpha^2)}\over -{1\over 1-\alpha^2}+{2\over p(1-3\alpha^2)+
p^2(1+\alpha^2)}}.
\ee
If $\Lambda$ is set equal to zero, this yields a relation between
$p,\alpha$. 
If one takes a negative value of $p$, the constant $C$ has to be set equal
to zero as mentioned above, and for finite $\alpha$, when $Q$ is non-zero,
one can get $f(r)$ decreasing with increase of $r$.
If $\alpha\to\infty$, one has $p\to(3\pm\sqrt{17})/2$
and $Q\to 0$. For the positive value of $p$, $f(r)$ has only the second 
piece 
\bb
f(r)=Cr^{-1}.
\ee
The equation for $c(r)$ is now satisfied {\it exactly} because $Q=0$.
The rotating black hole is a slowly rotating Kerr black hole: 
the corresponding non-rotating black hole [8] is a Schwarzschild solution.

\subsection{Third black hole}
The third non-rotating solution [8] is 
\bb
u(r)&=&r^{2\over 1+\alpha^2}\bigg(1
-{2M(1+\alpha^2)\over \alpha^2r}
+{(1+\alpha^2)^{2\alpha^2+1\over \alpha^2}\over (1-3\alpha^2)\alpha^2}
\Lambda Q^{2\over\alpha^2}r^{- 2\frac{1-\alpha^2}{1+\alpha^2}}
\bigg),\\
h(r)&=&r^{\frac{2\alpha^2}{1+\alpha^2}},\\
\phi(r)&=&\phi_0+\log r\frac{\alpha}{1+\alpha^2},,
\ee
with $\exp (-2\alpha\phi_0)= (1+\alpha^2)Q^2, b=-1/\alpha$.
%A single horizon is located at $r_h={2(1-\alpha^2)M \over (-1+2Q^2
%\exp 2\alpha\phi_0)\alpha^2}$, when this is positive.
The velocity for stable circular motion with $\theta=\pi/2$ turns out in 
the $\Lambda=0$ case to be $1/\alpha$ independent of $r$ for large $r$.
In case $\Lambda\neq 0$, it is $1/\alpha$ for $|\alpha|<1$ and 1 otherwise.

We are interested in finding the rotating version of this static solution,
that is to say, in solving the corresponding coupled equations for $f,c$. 
As in the previous cases, there exists the solution I ($c(r)=f(r)=0$) 
which is a trivial extension of the non-rotating solution to the equations
including rotation. The unusual solution
II ($c(r)=0,f(r)\propto h(r)$) where there is a change in the metric
without any change in the electromagnetic field also
exists as before. Again, this $f(r)$ increases with $r$. 
In addition, there are {\it asymptotic} solutions like
\bb
{\rm III:} &&c(r)= r^p,\nonumber\\
 &&f(r)=-(1+\alpha^2)Q^2pr^{p+\frac{1-\alpha^2}{1+\alpha^2}},
\ee
where $p,\alpha$ are elated by  $\alpha^2=\frac{p^2+p-4}{p(3-p)}< 1$.
As $\alpha^2\to 1-$, one value of $p$ goes to -1, the standard exponent.

For $\alpha^2=1$ and $\Lambda=0$, there is an exact solution
\bb
{\rm IV:} &&c(r)=r^2,\nonumber\\
&&f(r)=-4Q^2r^2+Cr,
\ee
with constant $C$.

Another exact solution exists for $Q=0$, which occurs for $\alpha^2\to\infty$.
\bb
{\rm V:} &&c(r)={\rm arbitrary},\nonumber\\
&&f(r)=Cr^2+Dr^{-1},
\ee
with $C,D$ constants. This is the form for a slowly rotating Kerr (anti-)
de Sitter black hole. The corresponding static case is the Schwarzschild
(anti-)de Sitter black hole obtained in the limit $\alpha^2\to\infty$.

\section{Discussion}
We have considered some non-rotating, charged,
dilaton black hole solutions for the action with a Liouville type
potential  $V(\phi)=2\Lambda e^{2b\phi}$ 
The theory involves the coupling parameter $\alpha$  besides
the potential parameter $b$: these were already
there in the non-rotating situation. The asymptotic behaviour of the
solutions is not flat, nor pure (anti-) de Sitter because of the
existence of the potential term which reduces to the cosmological
constant for vanishing dilaton field but is $\phi$- and hence
$r$- dependent in general. 
The solutions are not valid for general values of the parameters but hold
only for special relations between $b$ and $\alpha$, with isolated
singularities in $\alpha$.  The relations and the singularities are
different for the three classes as are the values of the $r$-independent
piece $\phi_0$ of the dilaton field. This 
leads in the three cases to different restrictions on
$\alpha$ if the potential is to vanish.

We have added an infinitesimal rotation measured by the parameter $a$. 
For small angular momentum the field equations lead to
the coupled differential equations (\ref{c}, \ref{f}) for $c(r), f(r)$,
for which we have presented several classes of solutions. 
In the case of the 
first black hole, solution III with $\alpha^2<1$ decreases 
for increasing $r$. Similar is the asymptotic solution V
for the second black hole when $p<0$ and $p+{1-\alpha^2\over 1+\alpha^2}<0$
and also when $p>0$ but $\alpha\to\infty$.
For the third black hole, III has a decreasing solution for negative $p$.
It would be of interest to find exact 
solutions with the angular momentum arbitrarily large, 
instead of being restricted to be small.

The absence of asymptotic flatness means that the metric
components do not converge to flat spacetime values for large $r$.
$h(r)$ increases with $r$ even in flat spacetime, and it is only the
the growth of $u(r)$ in the non-rotating black holes which may appear
surprising. But it has to be remembered that a growth occurs
also in the presence of a cosmological constant. The potential which is used
reduces to a cosmological constant for a vanishing dilaton field and can
therefore be expected to generate a growth in the more general case of
non-vanishing and indeed growing dilaton field. The electric potential
has the normal decreasing dependence on $r$ in the non-rotating case.
It is important to note that in spite of the increase with $r$ of $h,u,\phi$,
the terms in the action are finite and the curvature and
potential terms actually fall off with large $r$ [8].
This contrasts with (anti-) de Sitter solutions with a cosmological
constant where the curvature is constant.
The situation in the rotating case is more complicated: the new metric
function $f(r)$ increases with $r$ in some, though not in all of the solutions
presented here. In addition, the magnetic field associated with the
rotation of the charge is described by the function $c(r)$ 
which increases with $r$
in some solutions. The contributions to the action nevertheless fall off with 
increasing $r$ even here for the small values of $a$ used.

\bigskip

T. G. wishes to thank the Council of
Scientific and Industrial Research, India for financial support.

\end{document}